\documentclass[conference]{IEEEtran}

\IEEEoverridecommandlockouts

\usepackage{cite}
\usepackage{epsfig}
\usepackage{amsmath,amssymb}
\usepackage{mathtools}
\usepackage{bm} 
\usepackage[ruled,linesnumbered]{algorithm2e}
\usepackage{array}
\usepackage[caption=false,font=footnotesize]{subfig}
\usepackage{stfloats}
\usepackage{url}
\usepackage{balance}
\usepackage[colorlinks=true,linkcolor=black,citecolor=black]{hyperref}
\usepackage{xcolor}

\usepackage[shortcuts,acronym]{glossaries}
\newacronym{hrf}{HRF}{hemodynamic response function}
\newacronym{fus}{fUS}{functional ultrasound}
\newacronym{fmri}{fMRI}{functional magnetic resonance imaging}
\newacronym{snr}{SNR}{signal-to-noise ratio}
\newacronym{cpd}{CPD}{canonical polyadic decomposition} 
\newacronym{btd}{BTD}{block term decomposition}
\newacronym{corcondia}{CORCONDIA}{core consistency diagnostic}
\newacronym{eeg}{EEG}{electroencephalogram}
\newacronym{ecg}{ECG}{electrocardiogram}
\newacronym{iqr}{IQR}{interquartile range}
\newacronym{ml}{ML}{multilinear}
\newacronym{pcc}{PCC}{Pearson correlation coefficient}
\newacronym{sv}{SV}{singular value}
\newacronym{svd}{SVD}{singular value decomposition}
\newacronym{convcpd}{convCPD}{convolutional CPD}
\newacronym{tv}{TV}{total variation}
\newacronym{convbtd}{convBTD}{convolutional BTD}
\newacronym{lti}{LTI}{linear time-invariant}
\newacronym{roi}{ROI}{region of interest}
\newacronym{nvc}{NVC}{neurovascular coupling}
\newacronym{shiftcpd}{shiftCPD}{shift-invariant CPD}
\newacronym{pgd}{PGD}{projected gradient descent}
\newacronym{nls}{NLS}{nonlinear least squares}
\newacronym{bcd}{BCD}{block coordinate descent}

\newcommand{\reffig}[1]{Fig.~\ref{#1}}%

\newcommand{\ie}{i.e.,~}%
\newcommand{\eg}{e.g.,~}%

\newcommand{\mb}{\mathbf}
\DeclareMathOperator{\vecop}{vec}
\renewcommand{\vec}[1]{\vecop\!\left(#1\right)}

\newcommand{\Yt}{\mathcal{Y}}
\newcommand{\Xt}{\mathcal{X}}
\newcommand{\Ht}{\mathcal{H}}
\newcommand{\Et}{\mathcal{E}}

\newcommand{\tens}[1]{\mathcal{#1}}

\newcommand{\TR}{^\text{T}}    
\newcommand{\x}{\times}

\newcommand{\real}{\mathbb{R}}

\newcommand{\allr}{$r = 1, 2, \ldots, R$}
\newcommand{\allk}{$k = 1, 2, \ldots, K$,}
\newcommand{\matlab}{MATLAB\textsuperscript{\textregistered}}

\newcommand{\fus}{\gls{fus}}
\newcommand{\fmri}{\gls{fmri}}
\newcommand{\snr}{\gls{snr}}
\newcommand{\svd}{\gls{svd}}
\newcommand{\cpd}{\gls{cpd}}
\newcommand{\btd}{\gls{btd}}

\newcommand{\hrf}{\gls{hrf}}
\newcommand{\hrfs}{\glspl{hrf}}
\newcommand{\conv}{\gls{convcpd}}
\newcommand{\shift}{\gls{shiftcpd}}
\newcommand{\tv}{\gls{tv}}
\newcommand{\lti}{\gls{lti}}
\newcommand{\roi}{\gls{roi}}
\newcommand{\nvc}{\gls*{nvc}}
\newcommand{\iqr}{\gls{iqr}}
\newcommand{\pgd}{\gls{pgd}}
\newcommand{\nls}{\gls{nls}}
\newcommand{\bcd}{\gls{bcd}}
\DeclareMathOperator*{\rank}{rank}
\DeclareMathOperator{\mat}{mat}

\newcommand{\thetab}{\boldsymbol{\theta}}
\newcommand{\eps}{\mathrm{e}}

\begin{document}

\title{A Block-Term Decomposition Approach to Blind Multi-trial Functional Ultrasound Unmixing\thanks{The work of S.-E. Kotti has been supported by the TU Delft AI Labs programme. The work of E. Kofidis has been partly supported by the University of Piraeus Research Center.}}

\author{
\IEEEauthorblockN{
Sofia-Eirini Kotti$^{1}$,
Eleftherios Kofidis$^{2}$,
and Borb\'ala Hunyadi$^{1,3}$}
\IEEEauthorblockA{
$^{1}$\textit{Signal Processing Systems, EEMCS}, \textit{Delft University of Technology}, Delft, The Netherlands\\
$^{2}$\textit{Dept. of Statistics and Insurance Science}, \textit{University of Piraeus}, Piraeus, Greece \\
$^{3}$\textit{Mental Health and Neuroscience Research Institute, Maastricht University}, Maastricht, The Netherlands \\ 
Email: s.e.kotti@tudelft.nl, kofidis@unipi.gr, borbala.hunyadi@maastrichtuniversity.nl}
}

\maketitle

\begin{abstract}



Functional ultrasound (fUS) has emerged as a powerful neuroimaging modality due to its high resolution in both space and time, low cost and potential portability. Nevertheless, fUS signals provide only indirect observations of neuronal activity through the neurovascular coupling, and hence require the \emph{blind} separation of latent neuronal sources while also deconvolving their hemodynamic responses. In this work, we propose a data-driven convolutive block-term tensor decomposition-based model for multi-trial fUS measurements, where each source has a spatiotemporal representation comprising a low-rank spatial map and a piecewise-constant neuronal activation signal convolved with a trial- and source-dependent hemodynamic response function (HRF) with a physiologically plausible shape. We propose a constrained optimization framework for the model computation, which consists of alternating projected gradient descent iterations. Simulation results are reported that demonstrate accurate recovery of spatial maps and reliable estimation of activation temporal profiles across various noise levels, while confirming that HRF estimation remains the most challenging part of the problem.
\end{abstract}

\section{Introduction}
\label{sec:intro}

\Gls{fus}~\cite{mace2011} is an emerging neuroimaging modality that maps brain activity by measuring cerebral blood volume changes with power Doppler imaging. Similarly to \fmri, \fus\ signals provide an indirect observation of neuronal activity through the \nvc. Owing to its higher spatiotemporal resolution compared to \fmri, \fus\  has the potential to characterize the \nvc\ more accurately.  

\fus\  signals may contain contributions from task-related  sources, \ie driven by an experimental stimulus, or spontaneous neuronal activity, including physiological and acquisition-related artifacts. 
Separating and characterizing all underlying sources is of critical importance in \fus\ data analysis, since the activated regions and exact neuronal or hemodynamic contributions are generally unknown, even when approximate stimulus timings are available. 
Blind unmixing methods are well suited to address this challenge by estimating latent sources directly from the observed data, making assumptions only on the structure of the mixing mechanism.

In functional neuroimaging, the  \nvc\ is commonly modeled as a \lti\ system, whereby the measured signal is expressed as the convolution of neuronal activity with an (unknown) impulse response, referred to as the \hrf~\cite{Karahanoglu2013, Cherkaoui2021, Aydin2020}.
This convolutional framework underlies much of event-related \fmri\ research, where data-driven \hrf\ modeling, \eg in \cite{Makni2005, Sreenivasan2014, vanEyndhoven2021, Cherkaoui2021}, has demonstrated important advantages  over the use of fixed prior models, \eg the \textit{canonical} \hrf~\cite{Friston1998b}. 
In \fus, recent studies likewise adopt convolutive models for the mixtures of latent neuronal sources, \eg \cite{Erol2022, Kotti2024}. Further support for this modeling approach comes from simultaneous electrophysiology and \fus\ recordings, which have shown that \fus\ signals can be accurately predicted by filtering neuronal firing with an \hrf-like kernel~\cite{Nunez2022, Lambert2025}. Collectively, these findings motivate the joint estimation of neuronal activity and the \hrf\ from measured \fus\ data, as treating either quantity as known can bias the estimation of the other~\cite{Erol2022}. 

Given the intrinsically multidimensional structure of neuroimaging data, 
tensor models and methods~\cite{Sidiropoulos2017} have been widely adopted for their analysis, \eg \cite{Hunyadi2014, Chatzichristos2019, Hunyadi2017, Erol2022book}. By representing such data as tensors, low-rank tensor decompositions can exploit shared structure across dimensions and yield interpretable latent components~\cite{Morup2025}. Moreover, tensor decompositions readily accommodate constraints consistent with the underlying physiology.
Among these models, the \btd~\cite{Lathauwer2008b}, a generalization of the \cpd, is particularly well suited to the \fus\ scenario, as it allows each component to exhibit low-rank rather than strictly rank-one structure. This additional flexibility has similarly motivated the use of \btd\ in  \fmri\ applications~\cite{Chatzichristos2019}.

Blind unmixing of multivariate \fus\ time series, with \emph{joint} estimation of the \hrf\ and the spatiotemporal neuronal activity, was previously addressed in~\cite{Erol2022} via the rank-$(L_r,L_r,\cdot)$ \btd\ of the tensor of lagged measurement autocorrelation matrices, where a single, average time series per \roi\ was considered. A deterministic approach, not relying on statistical estimation, was proposed more recently in~\cite{Kotti2024}, where brain activity was extracted using the entirety of pixels and source components with low-rank spatial signatures were discovered through a two-step approach. First, a constrained rank-$(L_r,L_r,1)$ \btd\ was used to estimate the spatial maps, and subsequently, semi-blind deconvolution of the time signatures was employed to estimate the activation signals and the parametrized \hrf. However, in a two-step approach, errors in the estimated temporal signatures may propagate directly into the subsequent deconvolution stage. A \emph{joint} separation and deconvolution approach is expected to mitigate this effect by fitting the spatial maps, activation signals, and \hrf\ simultaneously, under shared constraints.

The present work considers a convolutional model of the \nvc\ for the case of \emph{multi-trial} \fus\ data.
Although commonly modeled as time-invariant, the \nvc\ can vary across trials due to changes in neuronal and physiological state~\cite{Larsson2016, Erol2024}, meaning that trial averaging may yield an oversimplified representation of the brain’s true response. Consistent with this observation, experimental studies in event-related \fmri\ have shown that the amplitude, latency, and dispersion of the \hrf\ can differ across repetitions of the same stimulus, motivating models with trial-specific \hrfs\ rather than a fixed response~\cite{Aguirre1998, Handwerker2004, Erol2024}. 
A related tensor-based work, the \shift\ model~\cite{Morup2008}, addresses this trial variability in \fmri\ by assuming that each component is characterized by a single temporal signature that undergoes a trial-dependent delay. 
This corresponds to fitlering with an impulse response with only one nonzero coefficient per trial. 
The \conv\ model~\cite{Morup2011} generalizes this formulation by assuming trial-dependent sparse impulse responses, thereby allowing each trial to contain a weighted sum of multiple delayed copies of the same temporal profile.

In this paper, we develop a model for multi-trial \fus\ data and an accompanying computation algorithm, with the following two contributions: a) we extend~\cite{Kotti2024} to the more realistic scenario of trial- and/or source-dependent \hrfs, while also merging the two steps of separation and deconvolution into one, and b) compared to~\cite{Morup2011}, we explicitly promote the low-rank structure of the spatial maps, through a rank-$(L_r,L_r,1)$ \btd\ instead of vectorizing the spatial maps, 
and we attach an \hrf\ shape and physiological interpretation to the trial- and source-dependent filters, instead of considering them sparse as in~\cite{Morup2011}. We further propose a constrained optimization framework for the model computation, which consists of alternating projected gradient descent iterations.
This is only the first step in a research project that aims to develop a complete tensor-based methodology 
for blind decomposition of \fus\ data. 




\subsection{Notation}

Vectors and matrices are denoted by lower- and upper-case boldface letters, respectively. We use calligraphic letters to designate higher-order tensors. For tensor $\Xt  \in \real^{I_1 \x \dots \x I_N}$, its mode-$n$ unfolding (matricization) is denoted $\mathbf{X}_{(n)} \in \mathbb{R}^{I_n \times \prod_{m \neq n} I_m}$. The superscript $\TR$ stands for transposition. The Frobenius and $\ell_1$ norms are respectively written as $\|\cdot\|_{\mathrm{F}}$ and $\|\cdot\|_1$. The symbols $\circ$ and $\ast$ are used to denote outer product and convolution, respectively. The vectorization operator, $\vec{\cdot}$, stacks the columns of a matrix into a single column vector. The converse operation of matricizing a vector is denoted by $\mat(\cdot)$.  Finally, $\real$ denotes the set of real numbers, and $\real_{+}$ denotes the set of nonnegative real numbers.

\section{System Model and Problem Statement}
\label{sec:problem}

Let us first recall the model proposed in~\cite[Eq.~(1)]{Kotti2024}, where the tensor of \fus\ measurements $\Yt \in \real ^{N_z\x N_x \x N_t}$, with $N_z$ and $N_x$ the numbers of pixels in the depth and width modes, respectively, and $N_t$ the number of time samples, is decomposable as in the following rank-$(L_r,L_r,1)$ \btd\ or otherwise known as LL1 tensor decomposition~\cite{Lathauwer2008b}
\begin{equation}
    \Yt \approx \sum_{r=1}^R \underbrace{(\mb{A}_r\mb{B}_{r}\TR)}_{\mb{U}_r} \circ (\mb{h}\ast\mb{z}_r),
    \label{eq:fUS}
\end{equation}
where $R$ is the number of components (sources) and $\mb{U}_r\in\mathbb{R}^{N_z\times N_x}$ is the spatial map for the $r$th source and is assumed to have a low rank $L_r < \min (N_z, N_x)$, consistent with the assumptions in \cite{Kotti2024,Chatzichristos2019} and the multisubject dictionary learning probabilistic atlas of \cite{Varoquaux2011}. 
Vector $\mb{z}_r\in\mathbb{R}^{N_t\times 1}$ represents the corresponding activation signal (or activation \textit{atom}), which is assumed to be piecewise constant~\cite{Karahanoglu2013, Cherkaoui2021, Kotti2024}, and $\mb{h}\in\mathbb{R}^{N_h\times 1}$ is the \hrf.  The convolution of the activation signal with the \hrf\ forms the mode-3 signature of the corresponding block term. In~\cite{Kotti2024}, the \hrf\ is assumed to be space- and source-invariant, and each activation signal may contain multiple trials (stimulus repetitions). In this paper, we instead consider multiple tensors $\Yt^{(k)}$ of the form~\eqref{eq:fUS}, with \allk\ one for each of $K$ trials. 
Here, the trial index may refer to repeated acquisitions of the same experimental protocol, obtained within the same recording session or possibly under different sessions, provided that the data are organized in a common format. Thus, the proposed formulation does not require explicit knowledge of the exact stimulus timings within each trial; rather, the temporal activation profiles are estimated from the multi-trial data.

\begin{figure}[!t]
    \centering    
    \includegraphics[width=0.4\textwidth]{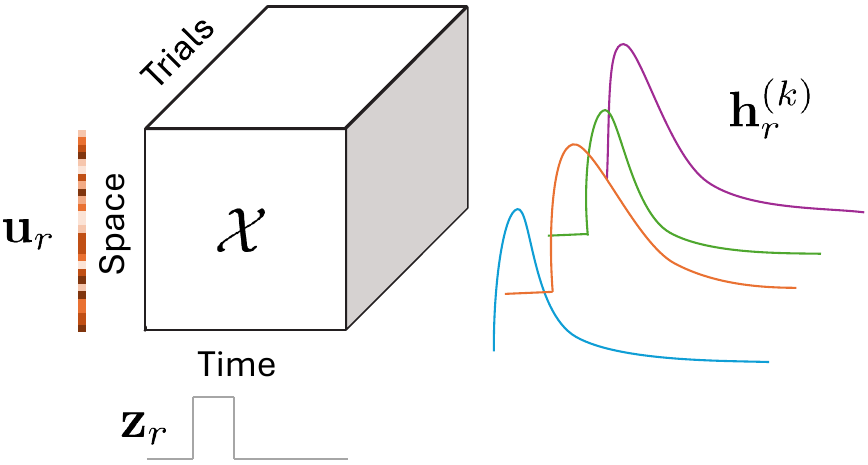}
    \caption{Proposed model for multi-trial \fus\ measurements.}
    \label{fig:convBTD}
\end{figure}

Relaxing the restrictive assumption of source-independent \hrfs, we can model each of these tensors as
\begin{equation}
    \Yt^{(k)} \approx \sum_{r=1}^R \mb{U}_r \circ (\mb{h}_{r}^{(k)}\ast\mb{z}_r), k=1,2,\ldots,K
    \label{eq:sfUS}
\end{equation}
Note that the source-dependent activation signals $\mb z_r$ and spatial maps $\mb{U}_r$ are shared across trials due to the repetition of the same stimuli (as is typical in functional neuroimaging). On the other hand, the source-dependent \hrfs~\cite{Hirano2011, Erol2022} can also be trial-dependent, and hence are denoted by $\mb{h}_{r}^{(k)} \in \real^{N_h \x 1}$, and are assumed to have common length $N_h < N_t$, without loss of generality.
If we stack all $\Yt^{(k)}$ into a four-dimensional tensor and subsequently reshape it to a three-dimensional tensor by combining the spatial dimensions into one, we arrive at tensor $\Xt \in \real^{N_{s} \x N_t \x K}$, with $N_{s}=N_zN_x$, whose frontal slices are the trial-specific matrices
\begin{equation}
    \Xt_{:,:,k} \approx  \sum_{r=1}^R \mb{u}_r \circ (\mb{h}_{r}^{(k)}\ast\mb{z}_r).
    \label{eq:X}
\end{equation}
In this expression, $\mb{u}_r=\vec{\mb{U}_r}\in\mathbb{R}^{N_{s}\times 1}$ and the assumption of low rank ($L_r$) for $\mb{U}_r$ translates to the requirement that $\mb{u}_r$ be a sum of a small number ($L_r$) of Kronecker products~\cite{Fu2019}. 
Viewed entry-wise, this tensor can be expressed as 
\begin{equation}
    \Xt_{i,j,k} \approx \sum_{r=1}^R \sum_{\tau=0}^{N_h - 1} u_{i,r}z_{j-\tau,r}h^{(k)}_{\tau,r}, \label{eq:convBTD}
\end{equation}
where $\mb z_r$ samples outside the interval $1, \dots, N_t$ are taken to be zero. The form in \eqref{eq:convBTD} resembles the \conv\ form~\cite[Eq.~(1)]{Morup2011}, albeit with the above-mentioned constraint on the $\mb{u}_r$ vectors. \conv~\cite{Morup2011} was introduced to accommodate trial variability in neuroimaging experiments: the authors assumed that each \cpd\ component is characterized by a spatial profile (here, $\mb u_r$) that is constant over trials, and an (unconstrained) temporal profile (here, $\mb z_r$), with each trial containing a weighted sum of time-shifted copies of this profile. This trial variability is encoded in the different filters (here, $\mb h^{(k)}_{r}$), which in~\cite{Morup2011} are forced to be sparse. The most distinctive characteristic of our model is that we impose low-rank structure on the spatial maps $\mb U_r$, which brings~\eqref{eq:X} closer to the constrained matrix factorization equivalent of LL1 proposed in~\cite{Fu2019} and will henceforth be referred to as \emph{convolutive BTD} (convBTD) to distinguish it from the \conv\ of~\cite{Morup2011}. Moreover, we here attach a \fus\ physiological interpretation to the involved quantities, as we did in~\cite{Kotti2024}: $\mb{z}_r$ stands for a piecewise-constant neuronal activation signal and $\mb{h}^{(k)}_{r}$ for the corresponding \hrf.  A schematic representation of the above is depicted in Fig.~\ref{fig:convBTD}. 

The problem of interest can then be stated as follows: given multi-trial \fus\ measurements organized in a third-order tensor $\Xt$, where each entry is modeled as a superposition in space of latent neuronal components convolved with unknown \hrfs, jointly estimate, for each of the $R$ underlying sources, the neuronal activation signal, $\mb{z}_r$, and the spatial map, $\mb{u}_r$, while also recovering the \hrf\, $\mb{h}_r^{(k)}$, for each source and each trial, in a data-driven manner. 
Blindly solving this unmixing problem is made feasible and facilitated through the explicit consideration of all \emph{a-priori} available information, incorporated here in the form of constraints. Thus, spatial maps $\mb{U}_r$, \allr\ are restricted to be nonnegative~\cite{Cherkaoui2021, Kotti2024} and of (low) rank $L_r < \min (N_z, N_x)$, in accordance with
~\cite{Varoquaux2011, Chatzichristos2019, Kotti2024}. 
Moreover, and in the spirit of~\cite{Fu2019}, nonnegativity will be directly imposed on $\mb{U}_r$ instead of so restricting its underlying factors as was done in~\cite{Kotti2024}.
For the activation signals $\mb z_r$, we assume piecewise constancy, in alignment with~\cite{Kotti2024, Cherkaoui2021, Karahanoglu2013}, and we adopt \tv\ regularization to force this property. Finally, we assume that the \hrfs\ respect the following well-known~\cite{Aydin2020, Kotti2023, Kotti2024} model
\begin{align} 
    h(t; \thetab) = \theta_1 (\Gamma (\theta_2))^{-1} \theta_3^{\theta_2} t^{\theta_2 - 1} \eps^{-\theta_3 t},
    \label{eq:hrf}
\end{align}
where $\thetab = [ \theta_1, \ \theta_2, \ \theta_3 ] \TR$ with $\theta_1,\theta_2,\theta_3\in\real_{+}$, and $\Gamma(\cdot)$ stands for the gamma function. This \hrf\ model is by definition nonnegative. Forcing the estimated \hrf\ to respect this model enhances the identifiability of the unknown parameters. Note that the present approach could also be applied in fMRI, after adapting the \hrf\ model accordingly. 

The above can be cast as the following optimization problem
\begin{eqnarray}
\lefteqn{\underset{\mb{U}, \mb{Z}, \Ht}{\min} \quad  F(\mb{U},\mb{Z},\Ht)=} \nonumber \\
& & \underbrace{\frac{1}{2} \sum_{i=1}^{N_{s}}\sum_{j=1}^{N_t}\sum_{k=1}^K \left(\tens{X}_{i,j,k}-\sum_{r=1}^R  \sum_{\tau=0}^{N_h-1} u_{i,r}  z_{j-\tau,r}h^{(k)}_{\tau,r} \right)^2}_{f(\mb{U},\mb{Z},\Ht)} \nonumber \\
& & + \lambda \underbrace{\sum_{r=1}^R \|\mb{D}\mb{z}_r\|_1}_{g(\mb{Z})} \label{eq:opt}
\end{eqnarray}
\begin{eqnarray*}
& \text{s.t.} & \mb{U}\geq \mb{0} \\
& & \rank(\mat(\mb u_r)) \leq L_r, r=1,2,\ldots,R \\
&  & \text{all\ } \mb{h}_{r}^{(k)}= \left[h(t_n; \thetab_r^{(k)})\right]_{n=1}^{N_h} \text{\ respect~\eqref{eq:hrf}}, 
\end{eqnarray*}
where $\mb{U} = [\,\mb{u}_1\ \cdots\ \mb{u}_R\,] \in \real^{N_s \times R}$, $\mb{Z} = [\,\mb{z}_1\ \cdots\ \mb{z}_R\,] \in \real^{N_t \times R}$, and the \hrfs\ are stacked into  tensor $\Ht \in \real^{K\times N_h\times R}$ with mode-2 unfolding given by $\mb{H}_{(2)} = [\, \mb{h}_1^{(1)} \ \cdots\ \mb{h}_1^{(K)} \ \cdots\ \mb{h}_R^{(1)} \ \cdots\ \mb{h}_R^{(K)} \,] \in \real^{N_h \times KR}$. The $N_t\times N_t$ matrix $\mb{D}$ is the discrete gradient operator. The regularization parameter $\lambda>0$ weighs the regularizer $g(\mb{Z})$ against the data fidelity term, $f(\mb{U},\mb{Z},\Ht)$.\footnote{It is natural to have the same weight for all $R$ activation signals since we expect their derivatives to have similar sparsity levels, as in visual experiments (e.g., \cite{Kotti2024}).} Our focus in this paper is on the solution and validation of the proposed model. We therefore make the simplifying assumption that $R$ and all $L_r$ are  known \emph{a priori}, an assumption that we relax in our future work. 

\noindent

\section{\btd-based Blind \fus\ Unmixing} 
\label{sec:algo}

Although the overall problem in~\eqref{eq:opt} is nonconvex, its multilinear structure enables a \bcd\ solution approach, in which the variables $\mb{U},\mb{Z},\Ht$ are updated alternately. Given the simplicity and low complexity of gradient descent, which also renders it an appropriate method for large-scale problems, plus the fact that it can quite easily incorporate constraints by simply projecting the result of each update on the feasible region~\cite{b1976}, we choose here to employ a \bcd\ procedure in which each sub-problem is addressed via \pgd\ iterations. Such an algorithm was also adopted in~\cite{Chen2022}, for LL1-based spectrum cartography, although it alternated between two matrices and did not involve TV regularization or convolution sums. 


To satisfy both the nonnegativity and low-rank constraints on \(\mathbf{U}\), we use an alternating projection procedure between the nonnegative orthant and the set of matrices whose matricized columns have rank at most \(L_r\). The projection onto the latter set is implemented by a rank-\(L_r\) truncated \svd\ of each matrix $\mat (\mb{u}_r) \in \real^{N_z\times N_x}$, while the projection onto the former is performed by the entry-wise application of $P_{+}(x)=\max(0,x)$. This procedure is motivated by composite property-mapping methods~\cite{Cadzow1988}, given that $\mathcal{N}=\real_{+}^{N_s\times R}$ is closed and convex and that the set $\mathcal{L}$ of matrices whose matricized columns have rank no greater than $L_r$, \allr\ is closed. 
Regarding the \hrfs, each column of $\mb{H}_{(2)}$ is projected on the \hrf\ manifold of~\eqref{eq:hrf}, through \nls\ curve fitting\footnote{In \matlab\ this can be done with the \texttt{lsqnonlin} function.}, which yields parameters $\thetab_r^{(k)} = \left[\begin{array}{ccc} \theta_{1, r}^{(k)} & \theta_{2, r}^{(k)} & \theta_{3, r}^{(k)}\end{array}\right]\TR$. Due to the nonconvexity of this \nls\ problem, we use multiple initializations and select the result with the best fit. 

To deal with the convex but non-smooth term $g(\mb{Z})$, there are several possible ways. We have selected the non-iterative method of~\cite{Condat2013}, which overcomes the non-smoothness of \tv\ via a proximal operator, $\mathrm{prox}_{\lambda g(\cdot)}$, and will be so referred to in the following. To ensure consistent regularization, the columns of $\mb{Z}$ are first normalized to unit $\ell_2$-norm, \ie $\|\mb{z}_r\|_2 = 1$.
The convolution of the \hrfs\ with the activation signals is performed in the frequency domain.
To this end, $\Ht$ is zero-padded along its temporal mode to the duration, $N_t$, of the activation signal.
The method is summarized as Algorithm~\ref{alg:convBTD}. 
\begin{algorithm}
\caption{convBTD}
\label{alg:convBTD}
        \KwData{$\Xt,\lambda,R,L_r,r=1,2,\ldots,R$}
        \KwResult{Estimates of $\mb{U}$, $\mb{Z}$, $\Ht$}
Initialize $\mb{U}$, $\mb{Z}$, $\Ht$\;
  \Repeat{convergence}{
	  Use \pgd\ with alternating projections onto $\mathcal{L}$ and $\mathcal{N}$ to update $\mb{U}$\;
    Use \pgd\ with projections onto \hrf\ manifold to update $\Ht$\;
    Use \pgd\ with $\mathrm{prox}_{\lambda g(\cdot)}$ to update $\mb{Z}$\;
}
\end{algorithm}

A couple of notes are in order here. One may solve for each block or instead accept an inexact solution. In our simulations, we have chosen to perform a (small) number of \pgd\ (i.e., inner) iterations. This was seen to improve the overall convergence speed. 
We have also implemented an adaptive step-size selection rule, in the vein of~\cite{b1976}. In each block update, the step size was initialized to the inverse of the local approximation of the Lipschitz constant for the corresponding gradient. The gradients and their Lipschitz constants were computed as follows. With tensor $\Et \in \real^{N_s\times N_t\times K}$ denoting the error $\Xt-\hat{\Xt}$, where $\Xt$ the measurement tensor and $\hat{\Xt}$ its convBTD reconstruction, we can write
\[
\mb X_{(1)} = \mb U (\mb Z \ast \Ht)_{(3)}+\mb{E}_{(1)}=\mb{U}\mb{V}\TR+\mb{E}_{(1)},
\]
where the convolution is understood along the common temporal mode of $\mb{Z}$ and $\Ht$, and, for convenience, we denote the $KN_t\times R$ matrix $(\mb Z \ast \Ht)_{(3)}\TR$ by $\mb{V}$. Then,
\[
\nabla_{\mb{U}}F = \left(\mb{U}\mb{V}\TR-\mb{X}_{(1)} \right)\mb{V}
\]
and
\[
\nabla_{\mb{U}}^2 F = \mb{V}\TR \mb{V}.
\]
Hence, the corresponding Lipschitz constant equals the square of the largest singular value of $\mb{V}$.
The gradients with respect to $\mb{Z}$ and $\Ht$ can be more conveniently expressed entry-wise. Indeed, we can write
\[
\frac{\partial f}{\partial Z_{j,r} } = -\sum_{\ell=1}^{N_{s}}\sum_{n=1}^K \sum_{\tau = 0}^{N_h - 1} U_{\ell,r}\Et_{\ell,j+\tau,n} \Ht_{n,\tau,r}
\]
and
\[
\frac{\partial f}{\partial \Ht_{k,\tau,r} } = -\sum_{\ell=1}^{N_{s}}\sum_{m=1}^{N_t}  U_{\ell,r} Z_{m-\tau, r} \Et_{\ell, m, k}.
\]
Their Lipschitz constants were approximated through their first-order divided differences. 
A more detailed description of Algorithm~\ref{alg:convBTD}, including its convergence analysis and the consideration of alternative possibilities, is deferred to an extended version of the present paper.

\section{Simulation Results}
\label{sec:sims}

We generated multiple datasets with $R=2$ task sources, each with an activation signal of length $N_t = 60$ samples, sampled at~4~Hz and consisting of a single nonzero block of random duration between~2 and~2.5 seconds. In our simulations, the activation signals were positive, but we did not enforce this in the optimization, similar to~\cite{Kotti2024}. In this way, neuronal deactivation in real data measurements can only be encoded in $\mb z_r$, given that the spatial maps and the \hrfs\ are assumed positive~\cite{Cherkaoui2021}. The spatial maps associated with the sources are of size $N_z = 20 \x N_x = 20$ pixels, as in~\cite{Kotti2024, Cherkaoui2021}, and have a common rank $L_r=L=5, r=1,2,\ldots,R$. Both the spatial maps and the activation signals were randomly chosen and allowed to overlap between the sources. We considered $K=4$ trials, so we generated $KR=8$~\hrfs\ that follow the model~\eqref{eq:hrf}. These \hrfs\ were kept fixed per component across all datasets to facilitate comparison of the results, and their parameters were constrained to realistic intervals as follows: 
$ 1 \leq \theta_{2, r} ^ {(k)} \leq 25$ and $0.5 \leq \theta_{3, r} ^ {(k)} \leq 25$, for all $r$ and $k$. These values cover the expected \hrf\ shapes in \fus\ data. Parameter $\theta_{1, r} ^ {(k)}$, which controls the \hrf\ amplitude,  is practically a free-running parameter that absorbs scaling during the updates, since the scaling ambiguity cannot be avoided in tensor decomposition without appropriate constraints. Therefore, we simply adjusted $\theta_{1, r} ^ {(k)}$  so that $\|\mb{h}_{r}^{(k)}\|_\infty = 1$ for all $r$ and $k$.
We generated the data tensors according to \eqref{eq:convBTD} and added white Gaussian noise to them at six \snr\ levels: -5, 0, 5, 10, 15, and 20~dB. For the \tv\ regularization, the \matlab\ implementation in~\cite{CondatCode} was used, as provided by the author of~\cite{Condat2013}.

\subsection{Training}

The training phase was used to tune the regularization parameter $\lambda$ for each \snr\ level. 
We generated~25 datasets, and selected $\lambda$ from a grid of~20 candidate values logarithmically spaced in the interval $[ 10^{-5}, 10^{-1}]$. For each dataset, the proposed algorithm was run with~20 random initializations, which were kept identical across \snr\ levels and $\lambda$ values to ensure a controlled comparison. The entries of the factor initializations were drawn independently from a standard Gaussian distribution, and the initial \hrfs\ and spatial maps were subsequently projected on $\mathbb{R}_{+}$ to satisfy the imposed nonnegativity constraints. 

Given that the ground truth quantities are available, the best run (combination of $\lambda$ and initialization) for each dataset was selected according to an oracle criterion, namely minimizing the aggregate estimation error of the recovered spatial maps, activation signals, and \hrfs. Specifically, we considered the sum of the component-averaged relative errors of the spatial maps and activation signals, and the worst-case relative error among the \hrf\ estimates over all components and trials.  The latter term was chosen to account for the fact that the \hrf\ estimation step is typically the most challenging part of the problem.
The stopping criterion was based on a relative change of the objective function below $10^{-9}$, while the maximum number of iterations was set to 50. Consequently, some runs may have been terminated before reaching the prescribed convergence tolerance.

\begin{figure*}
    \centering
    \begin{minipage}{\textwidth}
        \centering
        \subfloat[Spatial maps.]{\includegraphics[width=0.3\textwidth]{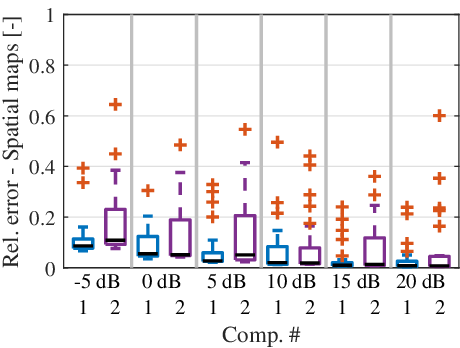}%
        \label{fig:spatial_maps_training}} \hfill
        \subfloat[Activation signals.]{\includegraphics[width=0.3\textwidth]{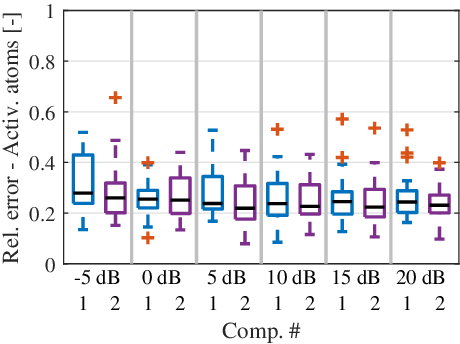}
        \label{fig:activ_atoms_training}} \hfill
        \subfloat[\hrfs.]{\includegraphics[width=0.3\textwidth]{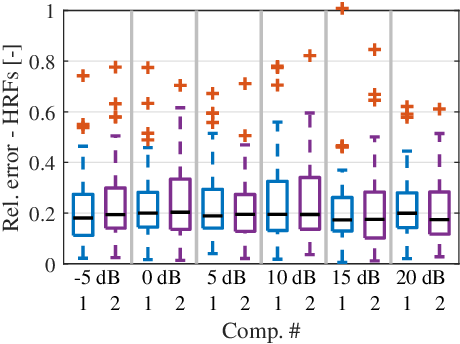}
        \label{fig:hrfs_training}} \hfill
        \caption{Boxplots of the relative error for the estimated quantities per \btd\ component during training at different \snr\ levels, defined as $\| \mb x - \hat{\mb x} \|_2 / \| \mb x \|_2$ for quantity $\mb x$ and its estimate $\hat{\mb x}$. Each boxplot in~(a) and~(b) includes 25~points (datasets), and, in~(c), 25 $\x$ 4 (datasets $\x$ number of trials). $\lambda$ differs per dataset and \snr. Colored boxes represent the \iqr, containing the middle $50\%$ of the data between the 25th ($q_1$)  and the 75th ($q_3$) percentile. Black lines indicate the median. Outliers (crosses) are points greater than $q_3 + 1.5 \cdot \text{IQR}$ or less than $q_1 - 1.5 \cdot \text{IQR}$. The whisker extends to the \textit{adjacent value}, the most extreme data point that is not an outlier.}
        \label{fig:training_est}
    \end{minipage}
\end{figure*}

The estimation results obtained during the training phase are shown in~\reffig{fig:training_est} in the form of boxplots of the relative errors of the recovered quantities. For each dataset, the plotted errors correspond to the selected $\lambda$/initialization pair according to the oracle criterion described above.
As shown in \reffig{fig:spatial_maps_training}, the spatial maps are accurately recovered over the considered \snr\ range. At $-5$ dB, the median relative error is approximately $0.1$ for both components, while the 75th percentile remains below approximately $0.2$. Both the median and the
\acrfull{iqr} decrease as the \snr\ increases, although a few outliers are observed at all \snr\ levels. The estimation of the activation signals, as shown in \reffig{fig:activ_atoms_training}, is moderately less accurate, with median errors of approximately $0.2$ and only few outliers. Finally, \reffig{fig:hrfs_training} indicates that the estimation of the \hrf\ is the most challenging part of the problem. Although the median relative errors are again approximately $0.2$, the 75th percentile and the outliers are substantially larger than for the spatial maps and activation signals. These results
suggest that the proposed algorithm can recover all three quantities with reasonable accuracy on average, while the performance remains dataset-dependent, particularly for the \hrf\ estimation.

Based on the above analysis, we selected the regularization parameter $\lambda$ per \snr\ level.
To reduce the influence of runs that may have been terminated before convergence, the value of $\lambda$ was selected using only datasets for which the relative errors for all three estimated quantities (spatial maps, activation signals, and \hrfs) did not exceed the corresponding 75$^\text{th}$ percentile. The final $\lambda$ values were then chosen as the median over the retained datasets:  $2.34 \times 10^{-2}$ at -5~dB, $8.86 \times 10^{-3}$ at 0 and 5~dB, and $5.46 \times 10^{-3}$ at 10, 15, and 20~dB.   

\subsection{Testing}

During the testing phase, the values of $\lambda$ selected during training were kept fixed for each \snr\ level. Performance was evaluated on 15 test datasets. For each dataset, the proposed algorithm was run with 30 random initializations, whose entries were drawn independently from a uniform distribution. 
The maximum number of iterations was increased to 70 in order to allow more runs to reach the prescribed convergence tolerance.

\begin{figure*}
    \centering
    \begin{minipage}{\textwidth}
        \centering
        \subfloat[Spatial maps.]{\includegraphics[width=0.3\textwidth]{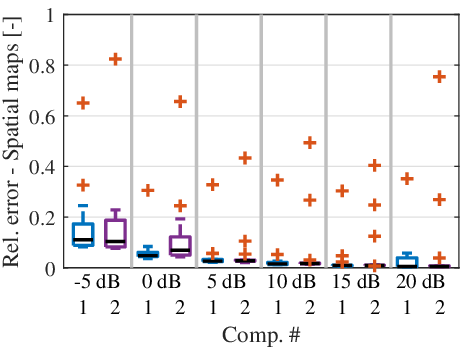}%
        \label{fig:spatial_maps_testing_est}} \hfill
        \subfloat[Activation signals.]{\includegraphics[width=0.3\textwidth]{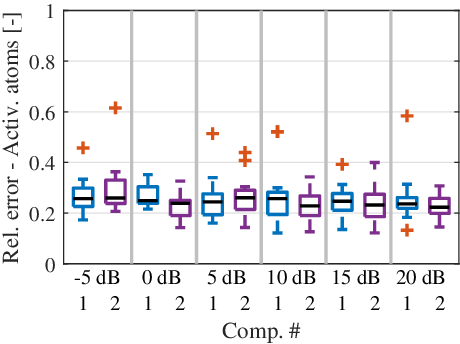}
        \label{fig:activ_atoms_testing_est}} \hfill
        \subfloat[\hrfs.]{\includegraphics[width=0.3\textwidth]{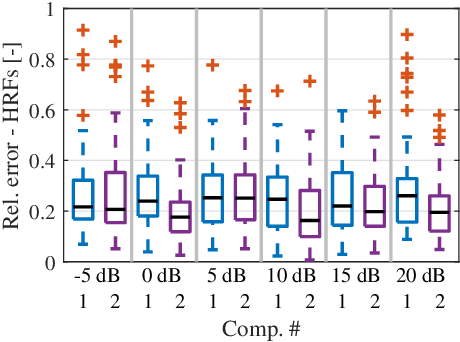}
        \label{fig:hrfs_testing_est}} \hfill
        \caption{As in~\reffig{fig:training_est}, with the initializations leading to best estimates being selected. Each boxplot in (a) and (b) includes 15 points (datasets), and in (c) 15 $\x$ 4 (datasets $\x$ number of trials). $\lambda$ differs per \snr, based on training results. Two outliers are excluded in (c) of value $\sim$1.3 at 5 and 15~dB.}
        \label{fig:testing_est}
    \end{minipage}
\end{figure*}

\begin{figure*}
    \centering
    \begin{minipage}{\textwidth}
        \centering
        \subfloat[Spatial maps.]{\includegraphics[width=0.3\textwidth]{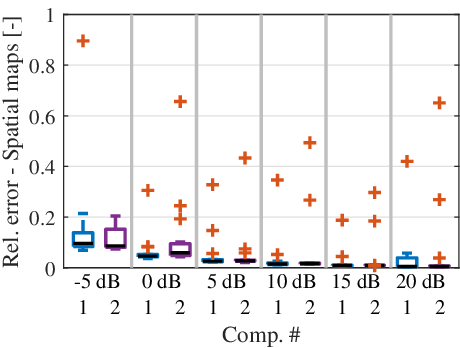}%
        \label{fig:spatial_maps_testing_err}} \hfill
        \subfloat[Activation signals.]{\includegraphics[width=0.3\textwidth]{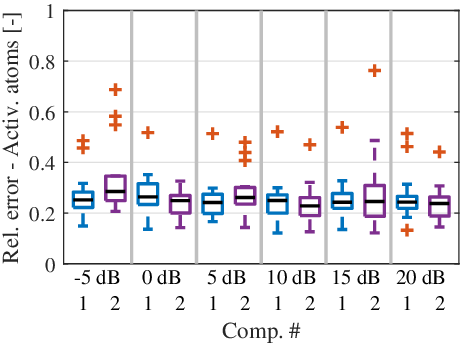}
        \label{fig:activ_atoms_testing_err}} \hfill
        \subfloat[\hrfs.]{\includegraphics[width=0.3\textwidth]{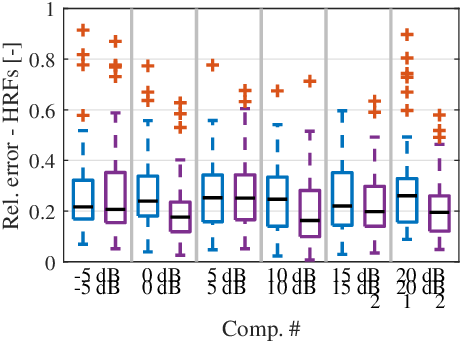}
        \label{fig:hrfs_testing_err}} \hfill
        \caption{As in~\reffig{fig:training_est}, with the initializations leading to the best data reconstruction being selected. Each boxplot in (a) and (b) includes 15~points (datasets), and, in (c), 15 $\x$ 4 (datasets $\x$ number of trials). $\lambda$ differs per \snr, based on training results. A few outliers are excluded: one in (a), of value $\sim$3 at -5~dB, and five in (c), of value $\sim$1.1--1.8, one at each of -5, 5, 10, 15, 20~dB. }
        \label{fig:testing_err}
    \end{minipage}
\end{figure*}

We report two sets of testing results. In~\reffig{fig:testing_est}, the best initialization is selected using the same oracle criterion as in the training phase, namely the lowest aggregate estimation error with respect to the ground truth. In~\reffig{fig:testing_err}, the initialization yielding the lowest data reconstruction error was selected. 
The latter criterion reflects a realistic selection strategy in a practical scenario where ground truth is unavailable. Overall, the estimation performance in~\reffig{fig:testing_err} is only marginally worse compared to~\reffig{fig:testing_est} in the median and \iqr\ sense, and mainly in the case of the \hrf\ estimation. This suggests that, under the assumption of additive Gaussian noise, selecting the initialization with the smallest reconstruction error provides a viable strategy for practical applications.

The spatial maps in \reffig{fig:spatial_maps_testing_err} are extracted almost perfectly for \snr\ at least 5~dB, with very few outliers. Even at lower \snr\ levels, the median relative error remains below~0.1, confirming the robustness observed during the training phase. The estimation of the activation signals in~\reffig{fig:activ_atoms_testing_err} exhibits a relatively narrow \iqr\ across all \snr\ levels. This behavior is consistent with using a fixed $\lambda$ for each \snr\ level, together with the fact that the derivatives of the simulated activation signals were of the same sparsity level. In contrast, the \hrf\ estimation errors in~\reffig{fig:hrfs_testing_err} show larger variability across datasets. This is consistent with previous observations~\cite{Kotti2024}, which report \hrf\ estimation as the most challenging aspect of this type of problem. The difficulty stems from the inherent ambiguity introduced by the convolution between the \hrfs\ and the corresponding activation signals. In the proposed multi-trial setting, we leverage the shared activation across trials to improve the \hrf\ estimation.

\begin{figure*}
    \centering
    \begin{minipage}{\textwidth}
        \centering
        \hspace{2cm} \subfloat[Relative estimation error for all quantities.]{\includegraphics[width=0.34\textwidth]{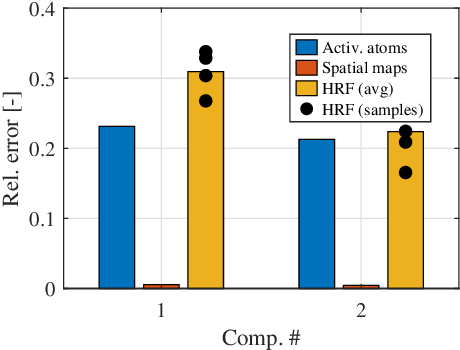}%
        \label{fig:set_35_errors}}
        \hspace{2.51cm}
        \subfloat[True and estimated activation signals.]{\includegraphics[width=0.34\textwidth]{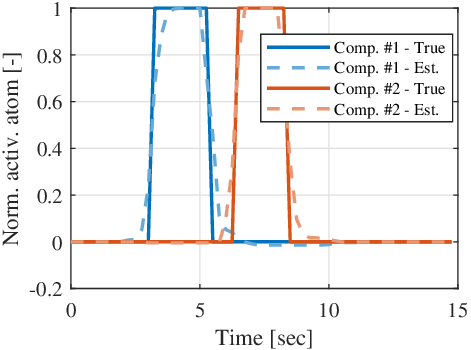}
        \label{fig:set_35_activ_atoms}}
    \end{minipage}
     \begin{minipage}{\textwidth}
        \centering
        \subfloat[True and estimated spatial map for component \#1.]{\includegraphics[width=0.6\textwidth]{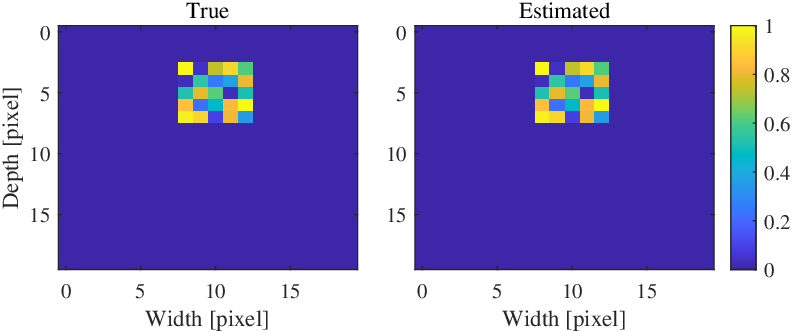}
        \label{fig:set_35_spatial_map_1}} \hfil
        \subfloat[True and estimated \hrfs\ for component \#1.]{\includegraphics[width=0.34\textwidth]{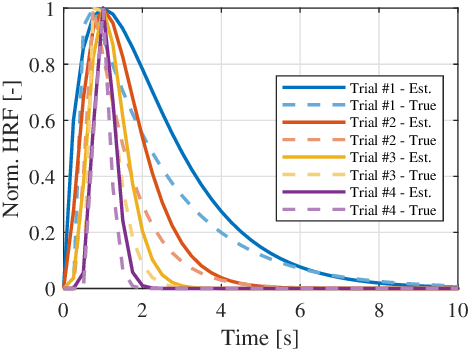}
        \label{fig:set_35_hrf_1}} \hfill
    \end{minipage}
    \begin{minipage}{\textwidth}
        \centering
        \subfloat[True and estimated spatial map for component \#2.]{\includegraphics[width=0.6\textwidth]{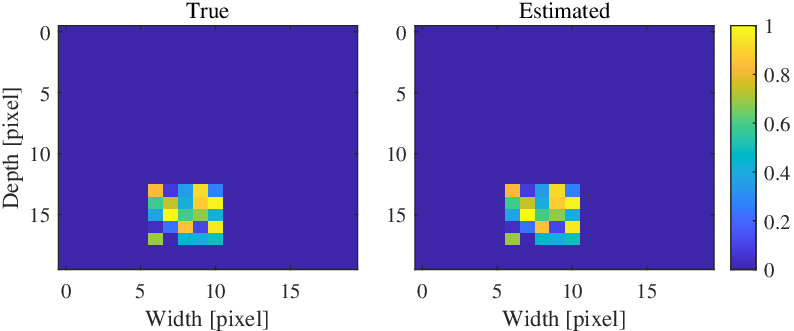}
        \label{fig:set_35_spatial_map_2}} \hfil
        \subfloat[True and estimated \hrfs\ for component \#2.]{\includegraphics[width=0.34\textwidth]{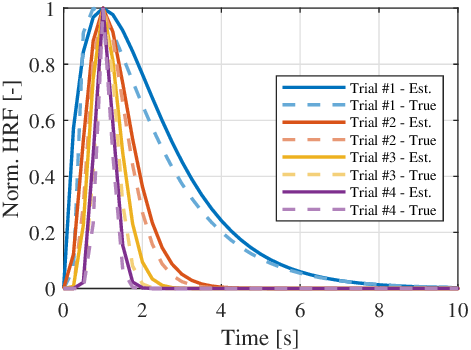}
        \label{fig:set_35_hrf_2}} \hfill
    \end{minipage}
    \caption{Estimation results for a single dataset.}
    \label{fig:set_35}
\end{figure*}

To further illustrate the behavior of the proposed method, \reffig{fig:set_35} shows the true and estimated quantities for a representative test dataset at 20~dB \snr, using the initialization with the minimum reconstruction error. This example is intended to complement the aggregate results in \reffig{fig:testing_err} by showing how high \hrf\ and activation signal estimation errors can occur.

The relative errors in \reffig{fig:set_35_errors} confirm that the spatial maps are accurately recovered for both components, whereas activation signals have relative errors around 0.2 for both components and the \hrfs\ between approximately 0.16 and 0.35. 
The spatial maps in \reffig{fig:set_35_spatial_map_1} and \reffig{fig:set_35_spatial_map_2} are recovered with almost perfect accuracy, both in terms of spatial support and intensity. As shown in \reffig{fig:set_35_activ_atoms}, the timing of the activation signals is captured almost perfectly. There are very small differences in the onset and offset times of the true and estimated signals; however, the estimated signals are smoother compared with the true ones that show sharper transitions. These differences account for the relative error of  roughly 0.2. The \hrf\ estimates in \reffig{fig:set_35_hrf_1} and \reffig{fig:set_35_hrf_2} are less accurate, particularly for the first component. The slightly more spread estimated activation signal leads to correspondingly more compressed \hrf\ estimates. This behavior illustrates the intrinsic ambiguity of the convolutional model: small changes in the temporal support of one factor can be compensated by opposite changes in the other, while still preserving a good reconstruction of the measured signal. The degree of this effect depends on the relative duration of the activation signals with respect to the \hrfs. Nevertheless, the estimated \hrfs\ still reproduce the main trial-dependent trend, namely the relative ordering of the \hrfs\ with respect to their temporal width. Overall, this example indicates that the proposed method can identify the activation timing and the associated low-rank spatial signatures, although accurate \hrf\ recovery remains affected by convolutional ambiguity. 

\section{Future Work}
\label{sec:conclusions}

Ongoing work in this context includes validating the proposed convBTD  framework on real multi-trial \fus\ recordings and investigating ways of (semi-)automatically estimating the number of sources, $R$, and their spatial ranks, $L_r$, via, \eg the \((L_r,L_r,1)\) core consistency diagnostic~\cite{Kotti2025}, appropriate regularization~\cite{Kofidis2022}, or Bayesian inference~\cite{Giampouras2022}. Beyond the single-subject setting, the proposed formulation could also be extended to multi-subject \fus\ data when subjects follow the same experimental protocol. In that case, the assumption of shared spatial maps would require anatomical co-registration of the \fus\ recordings into a common coordinate system. If such alignment is imperfect, subject-specific or partially shared spatial factors may be needed to account for inter-subject anatomical and functional variability. Finally, future work may include comparison with state-of-the-art convolutive independent component analysis and independent vector analysis (e.g., \cite{Dyrholm2007, Lehmann23}), as well as the investigation of alternative \hrf\ models, possibly also task-dictated~\cite{Moreno2021}.

\bibliographystyle{IEEEtran}

\balance

\bibliography{IEEEabrv,refs}
 
\end{document}